\documentclass[twocolumn,showpacs,preprintnumbers,amsmath,amssymb]{revtex4}

\input psfig.sty
\usepackage{graphicx}% Include figure files
\usepackage{dcolumn}% Align table columns on decimal point
\usepackage{bm}% bold math

\begin{document}
 %\def\be{\begin{equation}}
%\def\ee{\end{equation}}

 %\preprint{APS/123-QED}

\title{Constraining nonextensive statistics with
plasma oscillation data}

\author{R. Silva} \email{ rsilva@uern.br}
\affiliation{Departamento de Astronomia, Observat\'orio Nacional, 20921-400 Rio de Janeiro - RJ, Brasil}

\affiliation{Departamento de F\'{\i}sica, Universidade do Estado do Rio Grande do Norte, 59610-210, Mossor\'o - RN, Brasil}

\author{J. S. Alcaniz} \email{ alcaniz@on.br} 
\affiliation{Departamento de Astronomia, Observat\'orio Nacional, 20921-400 Rio de Janeiro - RJ, Brasil}

\author{J. A. S. Lima} \email{ limajas@astro.iag.usp.br} \affiliation{IAG, 
Universidade de S\~ao Paulo, 05508-900, S\~ao Paulo - SP, Brazil\\Departamento 
de F\'{\i}sica, Universidade Federal do Rio Grande do Norte, C.P. 1641, Natal - 
RN, 59072-970, Brasil}

\date{\today}

\begin{abstract}
We discuss experimental constraints on the free parameter of
the nonextensive kinetic theory from measurements of the thermal
dispersion relation in a collisionless plasma. For electrostatic
plane-wave propagation, we show through a statistical analysis
that a good agreement between theory and experiment is possible if
the allowed values of the $q$-parameter are restricted by $q=0.77
\pm 0.03$ at $95\%$ confidence level (or equivalently, $2-q =
1.23$, in the largely adopted convention for the entropy index
$q$). Such a result rules out (by a large statistical margin) the
standard Bohm-Gross dispersion relation which is derived assuming
that the stationary Maxwellian distribution ($q=1$) is the
unperturbed solution.
\end{abstract}
% PACS 
\maketitle

It is usually assumed that the particle velocity distribution in a
collisionless thermal plasma eventually relaxes to the standard
Maxwellian curve \cite{nkt,LD1}. However, the possibility of a
non-Gaussian behavior was established since the earlier
experiments in plasma physics \cite{L25}. More recently, Langmuir
probe measurements have also suggested that the isotropic
 component of the electron distribution resulting from inverse
 bremsstrahlung absorption in plasma is well described by a non
Maxwellian distribution \cite{liu}. In principle, such results may
be explained by a stationary distribution that emerges naturally
from the nonextensive statistical formalism proposed by Tsallis
\cite{const1}. Such an approach generalizes the standard
Boltzmann-Gibbs formalism through a new analytic form for the
entropy $S_q = k_{\rm{B}} (1 - \sum_{i}{p_{i}^{q}})/(q - 1)$,
where $k_{\rm{B}}$ is the standard Boltzmann constant, $p_i$ is
 the probability of the $i$th microstate, and $q$ is a parameter
 quantifying the degree of nonextensivity. This expression has been
 introduced in order to extend the applicability of statistical
mechanics to system with long range interactions and has the
standard Gibbs-Jaynes-Shannon entropy as a particular limiting
case ($q = 1$). The associated nonextensive kinetic theory has
also been considered in many different physical scenarios, ranging
from astrophysics to plasma physics (see \cite{Tsallisnet} for a
regularly updated bibliography). For example, it has been
successfully applied to two-dimensional Euler and drift turbulence
in a pure-electron plasma column \cite{BB96}, L\'evy-type
anomalous diffusion \cite{sd}, anomalous relaxation through
electron-phonon interaction \cite{ar}, ferrofluid-like systems
 \cite{fls}, plasma oscillations \cite{santos}, the solar neutrino
problem \cite{snp}, astrophysical systems \cite{nois2002},
among others. This new formalism has also shown to be endowed with
several interesting mathematical properties \cite{mathem}, with
the main theorems of the standard statistics admitting suitable
generalizations \cite{teoremas}. More recently, some authors have
argued that the value of the nonextensive parameter is predictable
for some simple physical systems, as for instance, in the case of
logistic maps. Their analyzes take into account the relation
between sensitivity to initial conditions and relaxation with the
q-generalized Lyapunov coefficient usually computed in terms of
the map parameters \cite{baldovin}. Theoretically, the situation
is not so neat for long-range Hamiltonian systems as it is for
maps. The present belief is that systems with short range
 interactions obeys the Boltzmann-Gibbs statistics and exponential
 relaxation dictates the approach to equilibrium. However, for more
 complex systems (including the case of plasmas), it is not
possible to determine apriori how the relaxation to a stationary
state takes place. In some cases, due to a formation of correlated
clusters in N-Body systems, a preequilibrium stage may happens
much earlier than the final Boltzmann-Gibbs equilibrium in such a
way that the overall effect can be described by a power-law
Tsallis distribution \cite{murilo}. The interesting feature of
this power-law function is that many models are analytically
tractable so that a detailed comparison with the Boltzmann-Gibbs
approach is immediate.
In this letter, we discuss new constraints on the $q$-parameter
associated to measurements of the dispersion relations for
electrostatic plasma oscillations. Through a statistical analysis
we estimate the value of the nonextensive parameter from the
existing experimental data. More precisely, by analyzing the data
set for thermal dispersion relation taken from Van Hoven \cite{VH}
we argue that a reasonable agreement between theory and experiment
is possible for values of the nonextensive parameter restricted by
$q=0.77 \pm 0.03$ at $95\%$ confidence level.
Some years ago, it was shown how Maxwell's derivation
\cite{maxwell} to the stationary velocity distribution function
could be extended to the nonextensive domain
\cite{rsilva,jaslima}. In the one-dimensional case, the new
distribution reads
\begin{equation} \label{q-function}
f_0(v_x)= A_q\left[1 - (q-1)\frac {mv_x^2}{2k_BT}\right]^{1 \over
{q-1}}.
\end{equation}
The above expression has been determined from two simple
requirements: (i) the isotropy of the velocity space, and (ii) a
nonextensive generalization of the Maxwell factorizability
 condition, that is, $F(v) \neq f(v_x)f(v_y)f(v_z)$\cite{rsilva}.
The quantity $A_q$ in the above equation denotes the
$q$-normalization constant and can be written, respectively, for
the intervals $0<q \leq 1$ and $q > 1$, by
\begin{equation}\label{A}
{A_{0 < q \leq 1}}={n\Gamma(\alpha)\over \Gamma({\alpha- {1 \over
2}})} \sqrt{m(1-q)\over 2\pi k_{B}T}
\end{equation}
and
 \begin{equation}\label{N1}
{A_{q > 1}} = n \left( {{1+q}\over 2} \right) {\Gamma({1\over
2}+\alpha)\over \Gamma(\alpha)} \sqrt{m(q-1)\over 2\pi k_{B}T},
\end{equation}
where $\alpha = 1/(q - 1)$ is a dimensionless number, $n$ is the
particle number density, $m$ is the mass of the particle and $T$
is the temperature. For $q>1$, the distribution (\ref{q-function})
exhibits a thermal cutoff on the maximum value allowed for the
velocity of the particles. However, we notice that the above
velocity distribution is parameterized with the nonextensive
 parameter shifted by $q \rightarrow 2-q$. This means that our
thermal cutoff at $q>1$ is equivalent to the cutoff condition,
$q<1$, usually adopted by other authors. As one may check, in the
extensive limit, $q=1$, the above distribution function reduce to
the standard Gaussian form \cite{santos,maxwell,rsilva}. Another
interesting result leading to a coherent nonextensive kinetic
theory is related to the transport equation. In this context, the
generalized Boltzmann's equation reads \cite{jaslima}
\begin{equation} \label{BQ}
{\partial f\over\partial t} + {\bf v}\cdot{\partial f\over\partial
{\bf r}}+{{\bf F}\over m}\cdot{\partial f\over
\partial{\bf v}}=C_q(f),
 \end{equation}
where the left-hand-side is the total time derivative of $f$ while
$C_q$ is the $q$-nonextensive collisional integral term
 responsible by changes in the distribution function $f$. In this
 approach, the stationary velocity $q$-distribution was obtained
from a slight generalization of the kinetic Boltzmann $H$-theorem.
The nonextensive ingredients follow simply by modifying the
molecular chaos hypothesis and by generalizing the local entropy
expression according to Tsallis' argument. One important related
result is that the proof of the $H_q$-theorem is possible (in
accordance to the second law) only if the nonextensive parameter
is positive definite, thereby fixing the lower bound assumed in
equation (2).
On the other hand, we know that high frequency vibrations in a
collisionless electronic plasma may kinetically be described in a
highly simplified manner, where electron-electron and electron-ion
collisions are unimportant, in such a way that the collisional
integral term in the Boltzmann equation may be neglected
 \cite{nkt,LD}. In the linear approach, the distribution function
of electrons is perturbed in first order, while the distribution
function of ions can be considered as an invariable quantity. If
$f_0(v)$ corresponds to the unperturbed homogeneous and
time-independent stationary distribution, the resulting particle
distribution function may be approximated as
\begin{equation}
f=f_0(v)+f_1({\bf r},{\bf v},t),\quad f_1\ll f_0,
\end{equation}
where $f_1$ is the corresponding perturbation in the distribution
function.
The dynamical behavior of the plasma can be described by a
combination of the linearized Vlasov and Poisson equations. One
obtains \cite{nkt,LD}
\begin{equation}
 \label{Boltz} \frac{\partial f_1}{\partial t} + {\bf
v}.\frac{\partial f_1}{\partial {\bf r}} = -{ e\over m}
{\nabla}\phi_1.\frac{\partial f_0}{\partial {\bf v}},
\end{equation}

\begin{equation}\label{poiss}
{\nabla^2}\phi_1 = -4\pi e{\int f_1({\bf r},{\bf v},t)d^3v},
\end{equation}
where $\phi_1({\bf r})$ is the first order correction in the
electrostatic potential and $e$ denotes the electronic charge. The
set of coupled equations (\ref{Boltz}) and (\ref{poiss}) may be
worked out through the simplified derivation for electrostatic
waves (longitudinal plasma waves), where the dispersion relation
can be calculated either by taking the constraint of null
 permitivitty \cite{LD1} or by using the mathematical techniques of
 integral transform originally developed by Landau \cite{LD}. In
this case, the solutions of the above equations can be written as
$f_1({\bf r},{\bf v}, t)=f_1({\bf v})\exp{i({\bf k}\cdot{\bf
r}-\omega t)}$ and $\phi_1({\bf r},t)=\phi\exp{i({\bf k}\cdot{\bf
r}-\omega t)}$ provided that $f_1$ and $\phi$ satisfy the
relations
\begin{equation}
 \label{relations} ({\bf k}\cdot{\bf v} - \omega)f_1({\bf v}) -
\phi{\bf k}.\frac{\partial f_0}{\partial{\bf v}} = 0
\end{equation}
and
 \begin{equation}
k^2\phi = 4\pi e\int f_1({\bf v})d^3v.
 \end{equation}
Finally, combining these expressions we find that the dispersion
relation between $\omega$ and ${\bf k}$ is given by
 \begin{equation} \label{BGt}
1 - \frac{4\pi e^2}{k^2 m}\int\frac{{\bf k}\cdot\,\partial
f_0/\partial{\bf v}}{{\bf k}\cdot{\bf v}-\omega} d^3v = 0.
\end{equation}
Now, by considering the limit case of small wave number $k\ll k_D$
(from now on the subindex $D$ stands for Debye quantities) we
expand the integrand of (\ref{BGt}) in power of $k$ and substitute
the nonextensive stationary distribution given by
 (\ref{q-function}) into the dispersion relation, one obtains (see
 reference \cite{santos} for more details)
\begin{equation}\label{BG}
 W^2 = 1 + 3(\lambda_Dk)^2\left({2\over 3q-1}\right),
\end{equation}
 where $W = \omega/\omega_0$, $\omega_0 = ({4\pi n e^2 \over
m})^{1/2}$ is the natural oscillation plasma frequency and
$\lambda_D=(k_B T/4\pi n e^2)^{1/2}$ is the electronic
Debye-H\"{u}ckel radius. As one may check, for the extensive limit
$q=1$, Eq. (\ref{BG}) reads
\begin{equation} \label{standard}
W^2 = 1 + 3(\lambda_Dk)^2,
\end{equation}

\begin{figure} [t]
%\vspace{.3in} 
\centerline{\psfig{figure=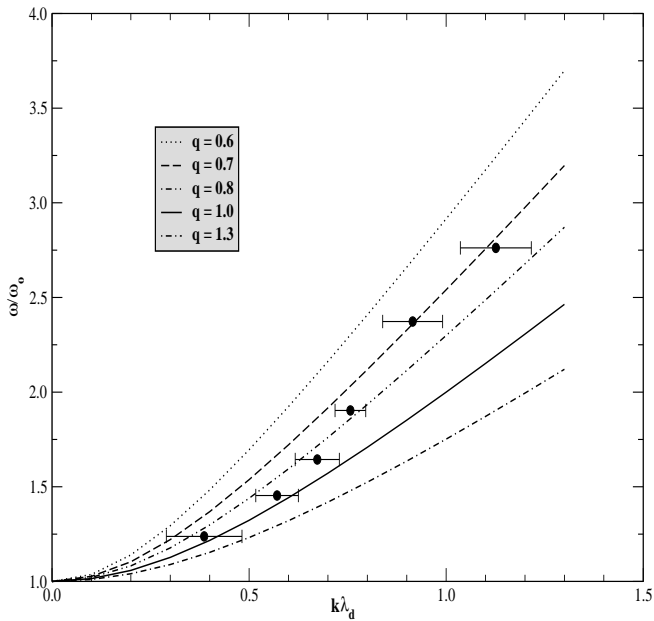,width=3.5truein,height=3.5truein,angle=-90}
\hskip 0.1in} 
\caption{Thermal dispersion relations for nonextensive velocity distribution. 
The selected values of $q$ is showed in the picture. Note that nonextensive 
distribution with $q <1$ is strongly suggested by these data.}
\end{figure}

which is the standard Bohm-Gross relation \cite{nkt}. In order to
check quantitatively the validity of this new approach, as given
by (\ref{BG}), we consider the experimental data set taken from
Van Hoven \cite{VH}. These data, originally composed by 40
experimental points, was distributed into 6 bins by using the
standard steps of data analysis (see, for instance,
\cite{kendall}). To find the confidence limits (c.l.) we use a
$\chi^{2}$ minimization neglecting the forbidden region by the
$H$-theorem ($q < 0$) \cite{jaslima}, i.e.,
\begin{equation}
 \chi^{2}(k \lambda_D, q) = \sum_{i=1}^{5}{\frac{\left[W(q) -
{W}_{oi} \right]^{2}}{\sigma_{i}^{2}}},
\end{equation}
where $W(q)$ is given by Eqs. (\ref{BG}) and (\ref{standard}) and
${W}_{oi}$ is the experimental values of the ratio
$\omega/\omega_0$ with errors $\sigma_{i}$ of the $i^{th}$ bin in
the sample.

\begin{figure} [t]
%\vspace{.3in} 
\centerline{\psfig{figure=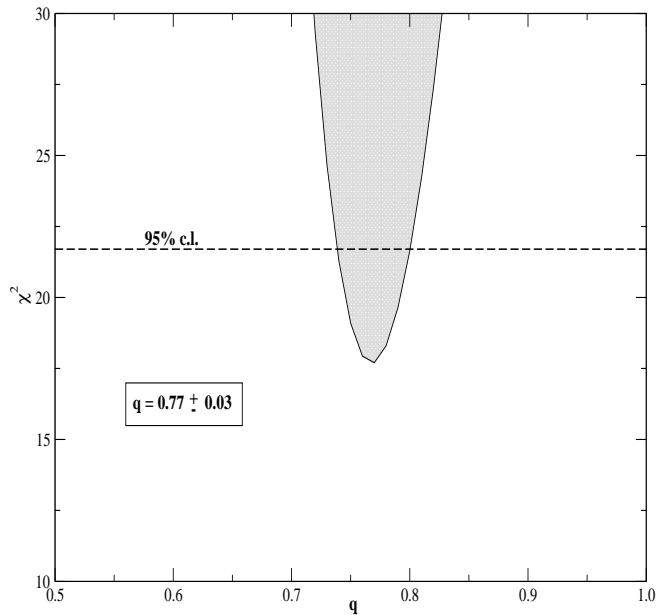,width=3.5truein,height=3.5truein,angle=-90}
\hskip 0.1in} 
\caption{$\chi^{2} - q$ plane provided by the data of Van Hoven [14]. The 
horizontal line indicates the 95$\%$ confidence limit for the nonextensive 
parameter. From this analysis we find $q=0.77 \pm 0.03$ (95\% c.l.).}
\end{figure}

In Fig. 1 we show the binned data of the ratio $W$ as a function
of dimensionless parameter $k \lambda_D$ for some selected values
of $q$. Thick solid line stands for the extensive Bohm-Gross
result based on the Maxwellian distribution ($q = 1$). As can be
seen, for values of $q> 1$ the curves increase less rapidly than
in the Maxwellian case and clearly depart from the experimental
points. In the complementary range, $0 < q < 1$, the curves
increase more rapidly than in the Maxwellian case, showing a
better agreement with the data. In Fig. 2 we display the result of
our statistical analysis in the $\chi^{2} - q$ plane. The horizontal dashed line corresponds to $95\%$ confidence level ($2\sigma$). For these data the peak of the likelihood is located
at $q = 0.77$ with the corresponding error interval of $\pm 0.03$.
Such a value suggests a power law behavior without thermal cutoff
for the allowed values of velocities. These results clearly show
that the new formalism discussed here may provide a better fit for
these experimental data than does the standard extensive approach
($q = 1$). Moreover, the present discussion reinforces the
interest for new experiments involving the dispersion relations in
a collisionless electronic plasma in order to check more
accurately the consistency of the results predicted by the
 nonextensive formalism.

In conclusion, we stress that the lack of a reasonable explanation
to the physical meaning of the $q$ parameter nowadays for
long-range Hamiltonian systems points naturally to the following
strategy: instead of paying attention to more formal results and
mathematical extensions, it seems more important to constrain its
value \emph{via} a number of different set of experiments (see,
 for instance, \cite{const1}), in order to determine more clearly
the reality of nonextensive effects. Actually, it has been shown
\cite{reis1} that the manganites (material that exhibit long-range
interactions and fractal geometry) can be studied through the
nonextensive statistic. In this regard, in contrast with
long-range interactions of manganites, the present study shows
that the role of long-range interactions does not seem to be a
fundamental ingredient for a statistical analysis of thermal
dispersion relation in plasma. In particular, we suspect that the
extended kinetic theory can also be linked with statistical
 correlations through the velocities (before and after of a
collision) of particles (see Refs. \cite{rsilva,jaslima}).
Finally, we have showed that the Bohm-Gross dispersion relation
which is a direct consequence of the Maxwellian distribution
($q=1$) is strongly deprived by the experimental data, being ruled
out by a considerably statistical margin. Our analysis indicates
that the best fit for these particular data set occurs for a
 nonextensive distribution without thermal cutoff with $q \simeq
0.77$ in our convention for the q-exponential (see equation (2)).
Note, however, that if the standard definition is considered (see
for instance \cite{jaslima,baldovin}) these data set constrains
the nonextensive parameter to the value $q=1.23$ which is
equivalent to a translation $q \rightarrow 2-q$.

\begin{acknowledgments}
This work is supported by the Conselho Nacional de Desenvolvimento
Cient\'{\i}fico e Tecnol\'{o}gico (CNPq - Brasil) and CNPq
(62.0053/01-1-PADCT III/Milenio). JSA is supported by CNPq (307860/2004-3)
\end{acknowledgments}

\end{document}